\documentstyle[prl,aps,epsf,floats,twocolumn]{revtex}
\newcommand{\beq}{\begin{equation}}
\newcommand{\eeq}{\end{equation}}

\def\eqa{\begin{eqnarray}}
\def\eea{\end{eqnarray}}
\begin{document}
\draft
\flushbottom
\twocolumn[
\hsize\textwidth\columnwidth\hsize\csname @twocolumnfalse\endcsname
\title{ Magnetic phenomena at and  near $\nu ={1\over 2}$ and ${1\over 4}$: theory, experiment and interpretation. }
\author{    R.Shankar  }
\address{
 Department of Physics, Yale
University, New Haven CT 06520}
\date{\today}
\maketitle
%\tightenlines
%\widetext
%\advance\leftskip by 57pt
%\advance\rightskip by 57pt

\begin{abstract}

I show that the hamiltonian theory of Composite Fermions   (CF) is
capable of yielding a unified description   in fair agreement with
  recent  experiments on  polarization $P$ and relaxation rate $1/T_1$ in  quantum Hall states
at filling $\nu = p/(2ps+1)$, at and near $\nu = {1\over 2}\
\mbox{and}\ {1 \over 4}$ ($s=1,2$) at zero and nonzero
temperatures. I show how rotational invariance and two
dimensionality can make the underlying interacting theory behave
like a free one in a limited context.
\end{abstract}
\vskip 1cm
\pacs{73.50.Jt, 05.30.-d, 74.20.-z}]

Recent  experiments  on  quantum Hall  systems have  focused  on
spin degrees of freedom. At $T=0$, critical fields $B^c$ at which
gapped states jump from one quantized value of polarization $P$ to
another\cite{du,yeh,kuk} and at which $P$ saturates for gapless
states have been measured\cite{kuk}. The temperature dependence of
$P$\cite{dem,mel}, and $1/T_1$\cite{dem} were measured for gapless
states. These data pose new constraints and challenges for
theorists. Why do  CF  appear to be strongly interacting, in that
it takes two substantially different masses $m_a$ and $m_p$ to
describe activation and polarization pheonomena; and yet
noninteracting, in that certain polarization phenomena can be fit
by  free fermions? Which mass should we use at  $T>0$?

The hamiltonian theory of CF, developed for small $q$\cite{gmrs}
and extended for all $q$ \cite{rs}, gives  a unified
 account of these experiments and answers the
question raised above. Once a {\em single} parameter characterizing
each sample is obtained by fitting the theory to one data point,
the values at other fields, tilts and temperatures are predicted.

This fermionic   theory is defined in first quantization by the
following equations. Readers not interested in the details may
safely go on
 Eqn. (\ref{zds})
:

\begin{eqnarray}
H &=&  \int {d^2q\over (2\pi)^2} {v(q)\over 2} \ e^{-{(ql)^2\over
2}}(\bar{\bar{\rho}}^p(q)\ \bar{\bar{\rho}}^p (-q) )
\   \ \  \label{1} \\
\bar{\bar{\rho}}^ p(q)&=& \bar{\bar{\rho }}(q) - c^2\
\bar{\bar{\chi}}(q)
\exp (-(ql)^2
/8ps) \\ \bar{\bar{\rho}}(q) &=& \! \sum_j \exp
(-i{\bf  q} \! \cdot \! ({\bf  r}_j\! - \! {l^2\over 1+c}\hat{\bf
z}\times {\bf \Pi}_j ))\\ \bar{\bar{\chi}}(q) &=& \! \sum_j \exp
(-i{\bf  q}\! \cdot  \!({\bf  r}_j \!+ \!{l^2\over c(1+c)}\hat{\bf
z}\times {\bf \Pi}_j ))\\ 0&=& \bar{\bar{\chi}} (q) |\mbox{Physical
State}\rangle \ \ \ \ \mbox{(constraint)} \\ c^2 &=& {2ps\over
2ps+1}\
\ \ \ \ \ \ \ \ \ \ \ \ \ \  \    l={1 \over
\sqrt{ eB}}\label{natural}
\end{eqnarray}
where
 $\Pi$ is the
 kinetic momentum of the CF in the  field $
B^* = {B \over 2ps+1}$,   $\hbar = c=1$, and the inoccuous spin
index has been suppressed. Referring to
Ref.\cite{gmrs,rs} for details, I merely state that these equations   provide an
operator realization of
 the idea\cite{jain,read1}  that in the lowest Landau Level (LL)  electrons bind to
$2s$-fold  vortices of charge $-c^2= -2ps/(2ps+1)$ (in electronic
units) to form composite fermions of charge
   $e^*  = e/(2ps+1)$.  Since  the CF  charge and dipole moment are built
in  at tree level, one expects a Hartree-Fock approximation to
be reliable.\cite{trans}

   I will use the  Zhang-Das Sarma \cite{zds} potential
\beq  v(q) = {2 \pi e^2 e^{-q\Lambda}\over \varepsilon q} \equiv
{2 \pi e^2 e^{-ql \lambda}\over \varepsilon q} \label{zds} \eeq
with $e^2/(\varepsilon lk_B)\simeq 50\sqrt{B(T)}{}^oK$. The value
of $\Lambda$ used to describe experiment will be discussed later.
The gaussian in Eqn. (\ref{1}), (absent in the small $q$ theory)
ensure that even for $\lambda =0$, (coulomb case) $q$ integrals
converge.

Note that $H=H_0+H_I$ is of a very unusual form. Consider  $H_0$,
the single particle part,  given by the diagonal terms in the
double sum     $\bar{\bar{\rho}}^ p(q) \cdot
\bar{\bar{\rho}}^ p(-q)$. In the simplest case $\nu
= {1
\over 2}, (s=1,\ p =\infty ,\ \ c=1 )$:
 \beq H_0
 =\sum_i 2\int {d^2q\over (2\pi )^2} \sin^2 \left[ {q\times
 k_i
 l^2\over 2} \right]v(q)e^{-(ql)^2/2}\label{ho}
 \eeq
 where $k$ is the particle momentum.
The CF kinetic energy originates from the interaction $v(q)$   as
it should, and is quadratic   only at small $k$. $H_I$ also has
nonstandard terms that appear with definite  relative strengths. At
$\nu \ne {1 \over 2}\ , {1\over4}$,  $H$ is even more complicated.
Fortunately, free-fermion states with filled LL and single
particle-hole excitations thereof, are its
 Hartree-Fock (HF)  eigenstates, and form the basis for the following computations.

Let us begin with gapped states at $T=0$, and assume unit area.
 Let  $|p-r,r\rangle$ denote a state with  $p-r$ spin-up and $r$ spin-down LL  of   CF's
 and hence a  polarization $P= (p-2r)/p$. Its energy
is  given in HF by
\beq E (p-r,r)   = \langle p-r,r|H|p-r,r \rangle .
\eeq

{\em The Zeeman energy, ignored here, can be trivially included.}
The critical field $B^c$ for the transition from  $r$ to $r+1$ is
found by equating $  E(p-r-1,r+1)-E(p,r+1)$ to the corresponding
Zeeman energy difference. All expressions can all be evaluated in
closed form.\cite{approach}

In the gapless case, $P$ is continuous and found by minimizing the
Zeeman energy plus the HF energy in a  state with Fermi momenta
$k_{\pm F}$ for the spin up/down species. Note that $
k_{+F}^{2}+k_{-F}^{2}=k_{F}^{2 }=4\pi n$, where $k_F$  is the
momentum of the fully polarized sea, and $n$ the total density. At
$T>0$, a HF calculation using these states gives $P$ and $1/T_1$ as
a function of $T$.

  Before turning to numbers,   I address and interpret a strange regularity in the $T=0$ results,
  already encountered by   Park and Jain\cite{parkjain}, namely that they
  can be fit by a theory of free fermions of
mass $m_p$   that  occupy LL with a gap $\Delta_p ={e B^*/ m_p}$. In
this case would have
\beq  E(p\! -\! r,r)-E(p\! -\! r\! -\!1,r\! +\! 1)=   {n(p-2r-1 )\over p}  \Delta_p  \label{gapdefine}
\eeq
 since    $(n/p)$ spin-up fermions of energy $(p-r-1+{1\over
2})\Delta_p$ drop to the  spin-down level with energy  $ (r+{1\over
2}) \Delta_p$.

Suppose we evaluate the left-hand-side of Eqn. (\ref{gapdefine}) in
the   HF approximation to  $H$ and {\em define}

\beq \Delta_{p}(r)^{def} = {p\over n} {
E(p-r,r)-E(p-r-1,r+1)\over p-2r -1}.\eeq

Given that  $H$ is not free, there is no reason why
$\Delta_p(r)^{def}$ should be  $r$-independent. But  it is very
nearly so. For example at $p=6, \lambda =1$,

\beq \Delta_p(0,1,2)^{def} = {e^2\over \varepsilon l}(0.00660, 0.00649,  0.00641)
\eeq
which describe $(6,0)\to (5,1)$, $(5,1)\to (4,2)$, and $(4,2)\to
(3,3)$. This was true for every  fraction and every  value of
$\lambda$ I looked at. Yet I knew that $H$  was definitely not free
since  the activation gap $\Delta_a$ to make a widely separated
particle hole pair differs from  $\Delta_p$ by factors like $2$ or
$4$ (depending on $\lambda$) and turning  off $H_I$ makes a
substantial difference.

Likewise at   $\nu ={1\over 2}$ and ${1\over 4}$  the
energy cost of flipping one spin could be fit to $(k_{+F}^{2}-k_{-F}^{2}) /2m_p$ where
 $m_p$ is independent of $P$ and
matches the value found at nearby gapped states. This too is
surprising given that ${\cal E}$, the HF energies of particles on
top either sea, does not even vary as $k^{2}_{\pm F}$! For example
at $\nu
={1\over 2}$
and $\lambda =1$
 \beq
{{\cal E}(k_{\pm F}) \over (e^2/\varepsilon l)} = a \left({k_{\pm
F} \over k_F}\right)^2 +b \left({k_{\pm F} \over k_F}\right)^4
\label{dispersion}\eeq where $a=.075$,    $b=-.030$.

I  will now explain these results for both gapped and gapless
states. Consider $E(S)$, the ground state energy density as a
function  of $S
= n\ P$, with up/down particles contributing $\pm 1$ to $S$. By
rotational invariance
 \beq E(S)= E(0) +
 {\alpha \over 2 } S^2 + {\cal O} (S^4)
 \eeq
Assume    ${\cal O} (S^4)$   terms are negligible. (We will see
that this does not correspond to a free-{\em fermion} assumption,
or in the gapless case, even  a $k^2$ kinetic energy.)

Consider now  the gapless case.
  When  $dn$  particles go from from  spin-down to spin-up,
 \begin{eqnarray}
 dE &=&{\alpha   } \ S\  dS = {\alpha   }\  S \ ( 2 \ dn)\\
 &=& \alpha {k_{+F}^{2}- k_{-F}^{2}\over 4\pi} (2 \   dn)
 \end{eqnarray}
using Luttinger's theorem on  the volumes of the Fermi seas.
  We see that $dE$  has precisely the form of the kinetic
 energy difference of particles of mass  $m_p$ given by
\beq
 {1 \over m_p} = {\alpha \over \pi}.
  \eeq
 Thus $m_p$ is really a susceptibility.
  Note that the free-field form of $dE$
  comes from  $E \simeq S^2$   {\em and}
   $d=2$: in $d=3$, we would have $dE/dn \simeq  S
 \simeq (k_{+F}^{3}- k_{-F}^{3}) $ which no one would intepret   as a
 difference of  kinetic energies.

 For gapped states,
  if $E\simeq S^2$ is  again assumed, the energies $E(p,r)$ will obey Eqn. (\ref{gapdefine})
  with an  $r$-independent
   $\Delta_p$. (Here one must replace $2S\ dS$ by the discrete
  expression $S(p-r,r)^2-S(p-r-1,r+1)^2.)$

To understand  the  smallness of the ${\cal O} \ (S^{4})$ term,
consider Eqn.(\ref{dispersion}). The cost of transferring a
particle from the top of one sea to the top of the other is

\begin{eqnarray}
{dE \over (e^2/\varepsilon l)}&=&{(a+b)\over
k_{F}^{2}}(k_{+F}^{2}-k_{-F}^{2})
\end{eqnarray}
using $k_{+F}^{4}-k_{-F}^{4}=(k_{+F}^{2}-k_{-F}^{2}) (k_{F}^{2}).$

{\em Thus the $k^4$ terms in ${\cal E}(k_{\pm})$ are not the cause
of the
 $S^4$ term.} However,    a small $ k^6$ term
 in ${\cal E}(k_{\pm})$, will generate small quartic terms in
$d E$ and $E(S)$.

To understand  why the $k^6$ term is so small, we turn to Eqn.
(\ref{ho}) for $H_0$. Expanding the $\sin^2$ in a series, we find
the $k^6$ term is down by a
 factor of at least 15 (50) relative to the $k^2$ term, at $\lambda
=0$  ($\lambda =1$), all the way up to $k=k_F$. Presumably this feature
 (and its counterpart in the gapped case)
persists in the HF approximation to  $H$ and keeps $E(S)$
essentially quadratic, which in turn mimics free-field behavior.

Now  I present the HF results for (the $r$-independent)  $m_p$ and $\Delta_p = eB^*/m_p$.  At and near $\nu = {1\over 2} \ \mbox{and} \ {1\over 4}$, for
  $.75 < \lambda < 2$, $m_p$  may be approximated by
\begin{eqnarray}
{1 \over m^{(2)}_{p}} &=& { e^2 l \over
\varepsilon }C_{p}^{(2)} (\lambda )\ \ \ \
 \ \ \ \  C_{p}^{(2)} (\lambda )  = {.087 \over \lambda^{7/4}} \label{mp2}\\
{1 \over m^{(4)}_{p}} &=& { e^2 l \over
\varepsilon }C_{p}^{(4)} (\lambda ) \ \ \ \
\ \ \ \
   C_{p}^{(4)} (\lambda )={.120 \over \lambda^{7/4}}\label{mp4}
\end{eqnarray}
 For fractions like $2/5$, not too close to $1/2$, I will use the actual $m_p$
  in comparing to experiment.

The transition $|p-r,r\rangle \to |p-r-1,r+1\rangle $ occurs when
\beq
 g{e\over 2m_e} {B_{\perp}^{c}  \over \cos \theta}
 =(p-2r-1)\Delta_p \label{tran}
             \eeq
where $g=.44$, $m_e$ is the electron mass, and  $\theta$ is the tilt.

In the gapless case
  the  total energy density $E_T(S)$,
 \beq
 E_T(S)={\alpha \over 2 } S^2 -g {e\over 2m_e} {B_{\perp} S \over \cos \theta}
\eeq
where  $\alpha  ={\pi / m_p}$, is minimized (for$P\le 1$) to give
 $P$:
\begin{eqnarray}
 P &=&  {.13 \sqrt{B_{\perp} }\lambda^{7/4}\over \cos
\theta}\label{22} \ \ \ \ \ \ \ \ \nu = {1 \over 2},\  \mbox{ B in Tesla} \label{27}\\ &=&\label{28} {.19
\sqrt{B_{\perp}}\lambda^{7/4}\over \cos \theta}\ \ \  \ \ \ \ \ \nu = {1 \over 4},\
\mbox{ B in Tesla}
\end{eqnarray}

At $T\ne 0$,  a HF calculation, in which the nonquadratic
dispersion in Eqn. (\ref{gapdefine}) and the corresponding density
of states play a central role,   yields $P$ and $1/T_1$.

We now compare to some experiments, starting with $\nu=1/2$ and $T>0$.
 Consider first Dementyev
{\em et al} \cite{dem} who find $P=.75$ for $B=B_{\perp}= 5.52 T$
at $300 \ mK$.  An  LDA calculation for the potential would give
$\lambda \simeq 1$ \cite{pj} for  this sample density. However,
since this number  does not include the effects of disorder, I have
chosen instead to match my HF results with the above data point,
(which gives $\lambda =1.75$) and see to what extent a {\em sole}
parameter $\lambda$, can
 describe  $P$ and $1/T_1$ for the
given sample at a given $B_{\perp}$, but various temperatures and
tilts.
 Since there does not exist a model, including disorder,  for  how $\lambda$
should vary with tilt  I include no such variation.

\begin{figure}
\epsfxsize=2.4in
\centerline{\epsffile{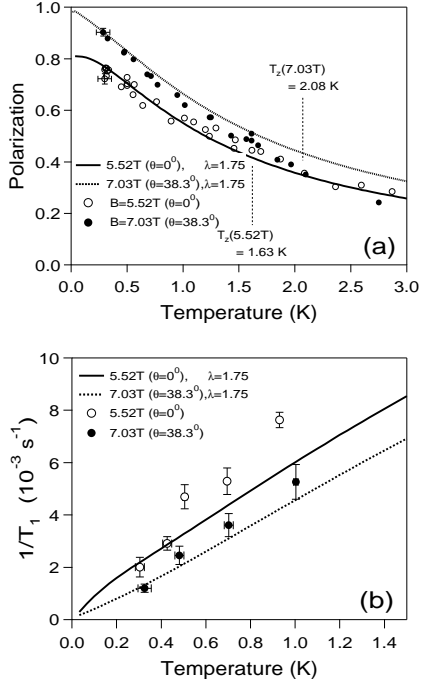}}
\vskip 0.15in
\caption{Comparison to the work of Dementyev {\em et al}.
  The value of $\lambda$ is fit to $P$ at $300 \ mK,\  B_{\perp}=5.52\ T$   and the  rest follows from the theory.
Notice the correlation between the curvature of
 $1/T_1$ and the limit of $P$ as
$T\to 0^{0}K$. }
\label{fig1}
\end{figure}

  Figure 1  compares the   HF calculation to the
  data.  Dementyev {\em at al} had pointed out that a two parameter fit
  (using a mass $m$ and interaction  $J$), led to four disjoint pairs of values
for these four curves. Given that $H$ is neither free nor of the
standard form ($p^2/2m +V (x) $) this is to be expected. By
contrast, a single $\lambda$  is able to describe the data here
since  $H$ has the right functional form. Given how the theory fits
the data (up to  the Fermi energy of $\simeq 1^o K$, beyond   which
point the size of the CF exceeds the spacing between them and the
CF concept breaks down) it is clear that changing the data point
used to fix $\lambda$ will be inconsequential.

  If $P$ were
computed from the LDA value  $\lambda
\simeq 1$, it would be down   by
15-50 $\%$ as $T$ drops from $1^o$K to $0^o$K.
 Thus  disorder and LL mixing are quite important,
and  at present  theory cannot describe experiments  {\em ab
initio}. Instead the present work establishes a phenomenological,
nontrivial  and  nonobvious fact that a single $\lambda$ parameter,
(like $g$ or $\varepsilon$)
 determined from one data point, can describe both $P$ and $1/T_1$
 for the given sample under a variety of conditions.
  That the fitted $\lambda$ is larger than the LDA value, makes sense, as
  both disorder and LL mixing will
 lower the gap and  raise $\lambda$.

 Consider next sample M280 of  Melinte {\em et al}
\cite{mel} which had $P=.76 $ at $.06^{o}K$ and $B=B_{\perp} = 7.1
T $, from which I deduced $\lambda =1.6$. Figure 2 compares my $T$
-dependence with data. The initial rise of $P$ with temperature
  was also  seen by  Chakroborthy and Pietlianean\cite{cp}.  As for the tilted sample, I do not
  understand the disagreement (not found in the Dementyev case).

\begin{figure}
\epsfxsize=2.4in
\centerline{\epsffile{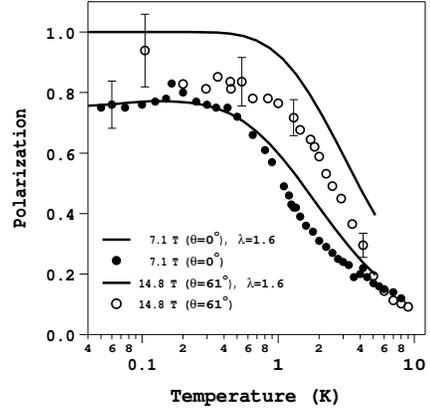}}
\vskip 0.15in
\caption{Comparison to  Melinte {\em et al}.
with $\lambda$  fit to $P$ at $60 \ mK,\  B_{\perp}=7.1\ T$. Some
typical error bars are  shown.}
\label{fig2}
\end{figure}

\begin{table}

\end{table}

Now consider Kukushkin {\em et al}  \cite{kuk} who  measure $P(B)$
at $0^{0}K$ by extrapolation. They vary both $B= B_{\perp}$ and the
density $n$ at each $\nu$ and observe, in addition to the spatially
uniform states of polarization labeled by $r$,    narrow steps of
intermediate polarization, probably      the inhomogeneous states
proposed by
  Murthy\cite{gm}.  In comparing to  theory, I will take the
   midpoints
  of these  steps as the transition points  between spatially
  uniform states.

I consider $B^c$'s at which the systems at $1/4,2/5,3/7,4/9, $ and
$1/2$ lose full polarization ($r=0$ for gapped cases, saturation
for the gapless cases) and, for $4/9$, also  the $r=1$ transition,
$(3,1)\to (2,2)$. I fit $\lambda$ to the  $\nu
=3/7$ transition
$(3,0)\to (2,1)$ at $B^c=4.5 T$. I obtain $\lambda_{3/7}=1.42$ on
solving Eqn.({\ref{tran}):

\beq g{e \over 2m_e}{  B^c\over (e^2/\varepsilon l)} = {2 \Delta_p
\left[ (3,0)\to(2,1)\right]\over (e^2/\varepsilon l)}={2  ( .0117)\over \lambda^{7/4}_{(3/7)}} \eeq

For  transitions  at other   $B_{\perp}$ and $n$,  I need the
corresponding $\lambda $'s. One can argue \cite{GMJain} that
$\Lambda \simeq n^{-1/3}$ so that $\lambda
\simeq n^{-1/3} B^{1/2} \simeq B^{1/6}\nu^{-1/3}$, and
\beq
\lambda_{\nu} = \lambda_{3/7} \left[ {B \over 4.5 }\right]^{1/6}\left[ {3 \over 7 \nu}\right]^{1/3} = .83 \ {B^{1/6}\over \nu^{1/3}}.\label{lambda}
\eeq

Given $\lambda$ one  finds $B^c$ using  Eqn. (\ref{tran}) for
gapped cases  and Eqns. ( \ref{27},\ \ref{28}) for saturation in
the gapless cases. The results are summarized in  Table I.

\vspace*{.2in}

\begin{tabular}{|c|c|c|c|c|} \hline \hline
$ \nu \ \ \  $ & comment &  \  $B^{c}$ (exp) &\  $B^{c}$ (theo) &\
\ $\nu B^c$ (exp)
\\ \hline
  4/9 & $(3,1)\to (2,2)$ & \ \ 2.7 T & \ \ 1.6 T & \ \  1.2\\
 2/5 & $(2,0)\to (1,1)$ &\ \ 3 T & \ \   2.65 T & \ \  1.2\\
1/4 & saturation &\ \   5.2 T &\ \  4.4 T &\ \  1.3\\ 3/7 & $(3,0)\to
(2,1)$ &\ \ 4.5 T&\ \   4.5 T & \ \    1.93\\ 4/9 & $(4,0)\to (3,1)$ &
\ \ 5.9 T & \ \  5.9 T & \ \   2.62\\ 1/2 & saturation&\ \  9.3 T&\ \
11.8 T&\
\  4.65 \\
\hline \hline
\end{tabular}

\vspace*{.2in}
Table I:  Critical fields based on a fit at $3/7$.The rows are
ordered by the last column which measures density.

\vspace*{.2in}

Note that in rows  above (below)  $3/7$, where I fit $\lambda$, the
predicted $B^c$'s are   lower (higher) than the observed values,
i.e.,  the actual $\lambda$'s are less (more) than what Eqn.
(\ref{lambda}) gives.
  This is consistent with the
expectation that
 interactions (neglected in Eqn. (\ref{lambda}))
   will increase
 the effective thickness with increased  density.
If I fit  to the $2/5$ point, I obtain similar numbers, with the
agreement worsening as we move  off in density from 2/5.

The work of Yeh {\em et al}\cite{yeh}  which involves an interplay
of $m_a$ and $m_p$,  and the work of Du {\em et al} \cite{du} which
deals with $\nu
>1$, fall outside the  present approach designed for $\nu <1/2$,
unless further assumptions  about the role of filled (electronic)
LL and particle-hole symmetry are made and justified.  A sequel
dealing with $m_a$ \cite{active}
 will  address these issues  and  provide details of the
  results merely sketched  here.  While the level crossings
  at the Fermi surface of Ref. \cite{yeh} can be understood in the interacting theory in the sense described here,
    crossings below are probably related  to  inhomogeneous states\cite{gm}.

In summary, the hamiltonian theory of CF \cite{gmrs,rs}  provides a
 quantitative description of several magnetic phenomena in FQH states at zero and nonzero temperatures.
A single parameter $\lambda$ extracted from one  data point,
provides a reasonable characterization  of each  sample at other
temperatures, tilts, fillings and $B_\perp$. This theory
accommodates enough interactions between CF to produce two distinct
masses $m_a$ and $m_p$, and yet a mechanism for simulating
free-fermion behavior in calculations of
    $B^c$ at $T=0$. At  $T\ne 0$, the interacting  theory,
   with its (nonquadratic)  dispersion relation and density of states, is  needed
   to correctly predict $P$ and $1/T_1$.

I  thank  J. Jain,  S. Sachdev and H. St\"{o}rmer for useful
discussions and S.Barrett and G. Murthy for teaching an old dog old
tricks. I am indebted to P. Khandelwal for his generous help in
numerically solving  the  HF equations and generating the figures.
I am pleased to acknowledge the support of grant NSF DMR98-00626.


\begin{thebibliography}{99}
\bibitem{trans} The constraint generates a gauge field. The effect of the
longitudinal part (renormaliation of   charge and dipole moment) is
built into $\bar{\bar{\rho}}^p$.
 The transverse part, which affects physics at the Fermi surface
at ultralow frequencies, is not included here, because of the gap,
 nonzero $T$, or the negligible impact it has on  total energies.
\bibitem{approach}  The  states are simple here while the
 operators for charge and $H$ are complicated. In
Jain's approach these simple wavefunctions get multiplied by
Jastrow  factors and projected into the LLL to yield complicated
wavefunctions in electronic coordinates, while $H$ is expressed in
terms
 of  the simple operator $\rho (q) = \sum_j e^{-iq \cdot r_j}$.
\end{thebibliography}
\end{document}